\documentclass[aip,jcp,reprint,amsmath,amssymb,superscriptaddress]{revtex4-2}

\usepackage{graphicx}
\usepackage{bm}
\usepackage{mathtools}
\usepackage{dcolumn}
\usepackage{hyperref}
\usepackage{xcolor}

\newcommand{\R}{\mathrm{R}}

\newcommand{\Sig}{\Sigma}

\begin{document}

\title{Voltage-Tunable Nonequilibrium Dispersion Interactions}

\author{Christine M. E. Little}
\author{Daniel S. Kosov}
\email{daniel.kosov@jcu.edu.au}
\affiliation{College of Science and Engineering, James Cook University, Townsville, QLD 4811, Australia}

\date{\today}

\begin{abstract}
We develop a nonequilibrium Green's function theory for dispersion interactions between two nanostructures, each an open quantum system driven into a nonequilibrium steady state by an applied bias voltage. Starting from the two-particle nonequilibrium Green's function, we derive a general expression for the interaction energy in terms of the polarisation propagators of the individual systems. The interaction energy admits a physically transparent decomposition into charge noise and charge dissipation contributions, providing a fluctuation-dissipation interpretation that generalises the equilibrium London picture. 
Model calculations for coupled molecular junctions demonstrate that the applied voltage can enhance the attractive dispersion interaction by nearly an order of magnitude relative to equilibrium. 
In thermal equilibrium, the dispersion interaction is universally attractive, irrespective of the specific form of the nanostructure Hamiltonians or their coupling to reservoirs. Out of equilibrium, we introduce a generalised Kubo-Martin-Schwinger ratio that parametrises the departure from detailed balance. We show that, in contrast to equilibrium, nonequilibrium conditions can lead to a repulsive dispersion interaction. Finally, we discuss the conditions under which population inversion in the electronic leads can drive a sign reversal of the dispersion interaction.
\end{abstract}

\maketitle

%=====================================================================
\section{Introduction}
\label{sec:intro}
%=====================================================================

Dispersion forces arise from correlated quantum fluctuations of charge densities on separate systems and constitute a fundamental component of intermolecular interactions.\cite{Stone2013,Israelachvili2011} These forces govern a wide range of phenomena from molecular adhesion\cite{Autumn2002}  and self-assembly to the stability of colloidal suspensions\cite{Israelachvili2011} and the physics of the Casimir effect.\cite{Klimchitskaya2009} In their equilibrium formulation, dispersion forces are universally attractive between polarisable bodies, a result that follows from the fluctuation-dissipation theorem.

The past two decades have seen rapid advances in nanoscale electronics, where individual molecules or quantum dots are contacted by metallic electrodes and driven far from equilibrium by applied bias voltages.\cite{Cuevas2017} In such molecular junctions, the fluctuation-dissipation theorem that underpins the standard theory of van der Waals forces ceases to hold. A natural question arises: how are dispersion forces modified when the interacting systems are each maintained in a nonequilibrium steady state (NESS)?
  A systematic theory of nonequilibrium dispersion interactions between current-carrying nanostructures has not yet been developed.

The question is equally pertinent to van der Waals heterostructures which are stacked two-dimensional layers such as graphene, transition-metal dichalcogenides, or hexagonal boron nitride, held together by dispersion forces.\cite{Novoselov2016}  In these devices each layer can be independently contacted and driven out of equilibrium, realising precisely the geometry of two capacitively coupled subsystems with no inter-system electron transfer that we analyse below. Coulomb drag experiments\cite{Gorbachev2012} already probe the capacitive interlayer coupling in this configuration;  whether the interlayer dispersion binding itself is modified by the current flow is an open question that the present theory is designed to address.

\begin{figure}[t]
\centering
\includegraphics[width=0.95\columnwidth, trim=3cm 2cm 3cm 2cm, clip]{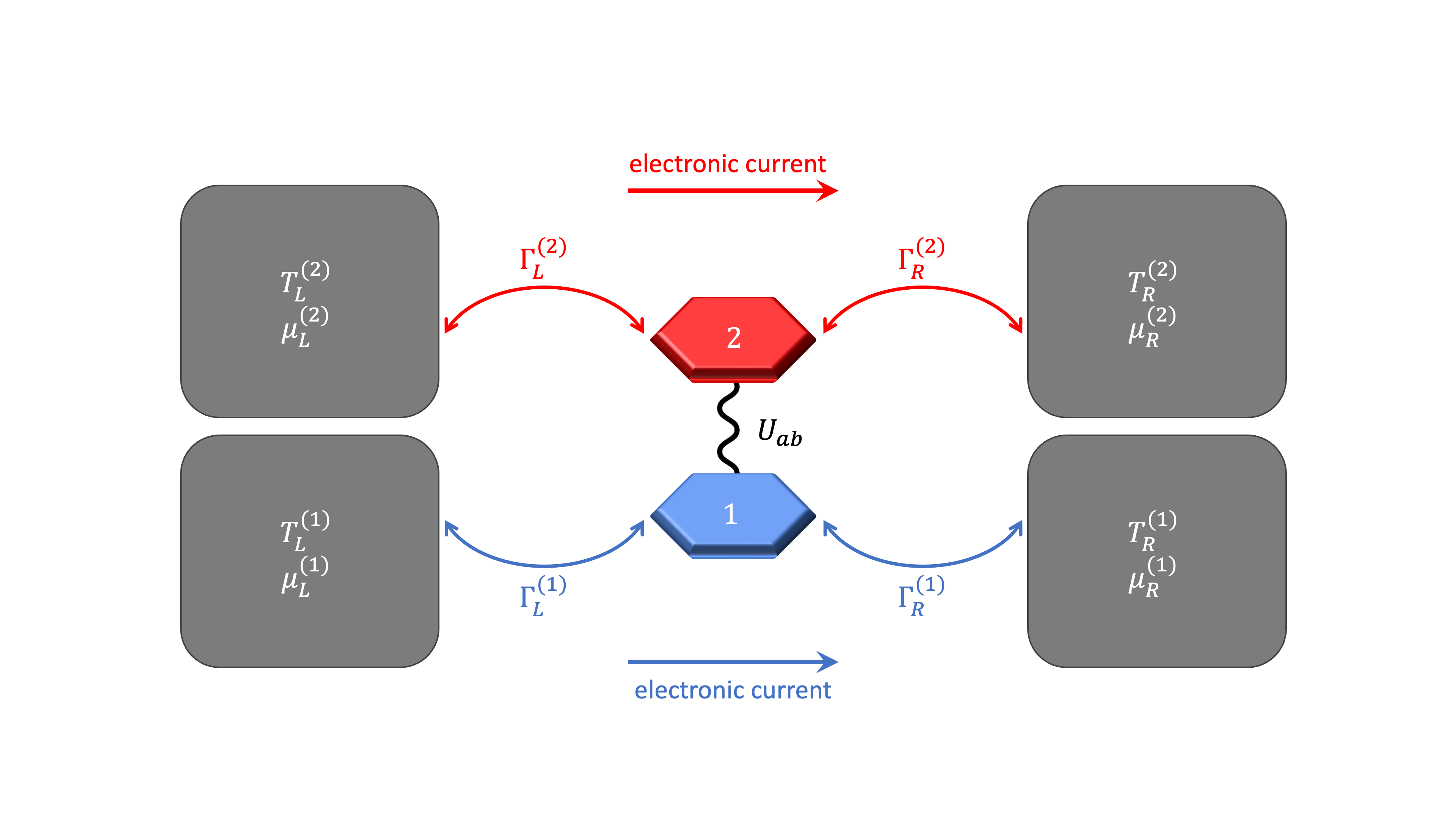}
\caption{Schematic of the setup. Two quantum nanostructures are each coupled to their own pair of electron reservoirs (leads). The two systems interact solely through the  Coulomb interaction, with no direct electron tunnelling between them.}
\label{fig:setup}
\end{figure}

In this paper, we present such a theory. Starting from the nonequilibrium Green's function (NEGF) formalism,\cite{Stefanucci2025,Haug2008} we derive a general expression for the dispersion interaction energy between two nanostructures that interact through a density-density Coulomb coupling, with no direct electron tunnelling between them (see Figure~\ref{fig:setup}). Each subsystem is individually coupled to its own pair of
electron reservoirs (leads), which drive it into a NESS. The interaction energy is obtained from the two-particle Green's function, with correlation self-energy evaluated at second order in the inter-system Coulomb coupling, yielding a formula expressed in terms of the polarisation propagators (particle-hole bubbles) of the individual subsystems.

We show that the dispersion interaction in NESSs can be expressed as a coupling between charge fluctuations and dissipation in the two subsystems. The interaction energy takes the form $E_{\mathrm{int}} \sim S_1 \chi_2 + \chi_1 S_2$, where $S(\omega)$ is the charge noise spectrum and $\chi(\omega)$ is the dissipative part of the density response. In equilibrium, these two quantities are locked together by the fluctuation-dissipation theorem; out of equilibrium, they become independent, and their interplay determines both the magnitude and the sign of the dispersion force.

We show that out of equilibrium, the interaction energy is controlled by a generalised Kubo-Martin-Schwinger (KMS) condition ratio $Z(\omega,V)$ that parametrises the departure from detailed balance. 
$Z(\omega,V)$ is the ratio of the rate at which the system
absorbs to the rate at which it emits a charge fluctuation of energy $\omega$ at voltage $V$.
The dispersion interaction is attractive when  $Z(\omega,V)> 1$, vanishes when $Z(\omega,V)=1$, and becomes repulsive under population inversion $ Z(\omega,V)< 1$.

We illustrate these general results with model calculations for two capacitively coupled single-level molecular junctions, using parameters representative of organic molecular conductors. The calculations demonstrate that realistic bias voltages can enhance the attractive dispersion interaction by nearly an order of magnitude relative to its equilibrium value. We show the emergence of repulsive dispersion interaction under  a population-inverted electronic distribution in the leads.

The paper is organised as follows. Sections~\ref{sec:setup} - \ref{sec:noise_response} develop the general theory, including the derivation of the interaction energy from the two-particle nonequilibrium Green's function, the second Born self-energy, and the noise-response decomposition. Section~\ref{sec:equilibrium} proves the equilibrium attractiveness of the dispersion force. Section~\ref{sec:noneqKMS} analyses the nonequilibrium case, introducing the generalised KMS condition. Section~\ref{sec:model} presents model calculations for capacitively coupled molecular junctions. Section~\ref{sec:discussion} discusses the physical mechanisms, conditions for repulsion, and connections to experiment. Section~\ref{sec:conclusions} summarises the conclusions.

%=====================================================================
\section{Theory}
\label{sec:theory}
%=====================================================================

\subsection{Setup}
\label{sec:setup}

We consider two quantum nanostructures (labelled 1 and 2), each connected to its own pair of electron reservoirs (leads) held at different chemical potentials and/or temperatures, establishing a bias voltage and/or temperature gradient across the systems.  The total Hamiltonian of the combined system is
\begin{equation}\label{eq:Htotal}
  H_{total} = H_1 + H_2 + H_{\mathrm{leads}} + H_{\mathrm{tun}} + H_{\mathrm{int}} \,.
\end{equation}
The first two terms describe the isolated nanostructures:
\begin{align}\label{eq:Hdot}
  H_1 &= \sum_a \epsilon_a d_{a}^{\dagger} d_{a} \,,
  \\
  H_2 &= \sum_b \epsilon_b  d_{b}^{\dagger} d_{b}  \,,
\end{align}
where $\epsilon_a$ are the single-particle orbital energies and $d_a^{\dagger}$ ($d_a$) are the fermionic creation (annihilation) operators for orbital $a$ on subsystem $1$. The Hamiltonian $H_2$ analogously describes nanostructure $2$. In the following, indices $a, a_1, \ldots$ are reserved for orbitals on subsystem $1$, while $b, b_1, \ldots$ label orbitals on subsystem $2$.

The leads are modelled as non-interacting electron reservoirs:
\begin{equation}\label{eq:Hleads}
  H_{\mathrm{leads}} = \sum_{\nu=1,2}\;\sum_{\alpha=L,R}\;\sum_k
  \varepsilon_{k\alpha \nu}\, c_{k\alpha \nu}^{\dagger}\, c_{k\alpha\nu}\,,
\end{equation}
where $c_{k\alpha\nu}^{\dagger}$ creates an electron in state $k$ of lead $\alpha \in \{L,R\}$ attached to subsystem $\nu$.

The tunnelling Hamiltonian couples each nanostructure to its own leads:
\begin{align}\label{eq:Htun}
  H_{\mathrm{tun}} &= \sum_{\alpha=L,R}\;\sum_{k,a}
  \bigl( t_{k\alpha,a} \, c_{k\alpha1}^{\dagger}\, d_{a} + \mathrm{H.c.} \bigr) 
  \\
  &+
  \sum_{\alpha=L,R}\;\sum_{k,b}
  \bigl( t_{k\alpha,b} \, c_{k\alpha2}^{\dagger}\, d_{b} + \mathrm{H.c.} \bigr)
  \,.
\end{align}
where $t_{k\alpha,a}$ is the tunnelling amplitude between state $k$ of lead $\alpha$ and orbital $a$ of subsystem $1$. Likewise, $t_{k\alpha,b}$ denotes the tunnelling amplitude between state $k$ of lead $\alpha$ and orbital $b$ of subsystem $2$.

There is no direct electron tunnelling between the two nanostructures; they interact only through the density-density Coulomb coupling
\begin{equation}\label{eq:Hint}
  H_{\mathrm{int}}
  = \sum_{a,b} U_{ab}\,\hat{n}_a\,\hat{n}_b \,,
\end{equation}
where $\{a\}$ and $\{b\}$ are orbital indices on systems $1$ and $2$, respectively, $\hat{n}_a = d_a^\dagger d_a$ and $\hat{n}_b = d_b^\dagger d_b$ are the occupation number operators, and the coupling matrix $U_{ab} = U_{ba}$ is real and symmetric.

The leads attached to nanostructure $1$ are assumed to be characterised by chemical potentials $\mu^{(1)}_{\alpha}$ and temperatures $T^{(1)}_{\alpha}$; likewise, the leads connected to nanostructure $2$ are described by chemical potentials $\mu^{(2)}_{\alpha}$ and temperatures $T^{(2)}_{\alpha}$.

\subsection{Interaction energy from the two-particle Green’s function}
\label{sec:Eint_deriv}

We derive an expression for the interaction energy by writing the correlation self-energy in terms of the two-particle Green’s function on the Keldysh contour.\cite{Schlunzen_2020}
The single-particle Green’s function is defined as
\begin{equation}
G_{pq}(\tau_1, \tau_2) 
= -i 
\langle    \hat T_C ( d_{p}(\tau_1)  d^\dag_{q}(\tau_2) ) \rangle\,,
\end{equation}
and the two-particle Green’s function as
\begin{equation}
G^{(2)}_{pqrs}(\tau_1,\tau_2,\tau_3,\tau_4) 
= (-i)^2 
\langle    \hat T_C (  d_{p}(\tau_1)  d_{q}(\tau_2)     d^\dag_{r}(\tau_3)    d^\dag_{s}(\tau_4)  ) \rangle\,.
\end{equation}
Here, $\tau$ denotes time on the Keldysh contour, $\hat T_C$ is the contour-time ordering operator, and the creation and annihilation operators are in the Heisenberg picture. The indices $p,q,r$ and $s$ label orbitals on either nanostructure $1$ or $2$.

The correlation self-energy is defined through the following integral equation
\begin{multline}\label{eq:SD_instant}
  -i \sum_b \; U_{ab}\, G^{(2)}_{bapb}(\tau_1,\tau_1, \tau_2, \tau_1^+)
  \\
  =\sum_q \int_C d\tau_3 \Sig_{aq}(\tau_1, \tau_3)\,G_{qp}(\tau_3, \tau_2) \,,
\end{multline}
where $\tau^+ =\tau +\eta$ is shifted with infinitesimal contour time $\eta$ and ${\Sig}$ is the correlation self-energy arising from the inter-system coupling $H_{\mathrm{int}}$.
We now select the contour-time arguments to extract the interaction energy. First, we place $\tau_1 = t_{1+}$ on the forward branch and $\tau_2 = t_{1-}$ on the backward branch (both at the same physical time $t_1$). Next, setting $p = a$ and summing  over orbitals ($\sum_a$) for nanostructure $1$, we observe that the left-hand side of (\ref{eq:SD_instant}) becomes  $i$ times the expectation value of $H_{\mathrm{int}}$ at time $t_1$.  Therefore,
the interaction energy at time $t_1$ becomes
\begin{equation}\label{eq:Eint_contour}
  E_{\mathrm{int}}(t_1)
  = -i \int_C d\tau_3\;
  \sum_{aq} \bigl[ \Sig_{aq} (t_{1+},\tau_3)\, G_{qa}(\tau_3,t_{1-}) \bigr] \,.
\end{equation}
We now convert from contour to real time.  Splitting the contour integral over $\tau_3$ into forward and backward branches,
one obtains
\begin{multline}
  E_{\mathrm{int}}(t_1)
  = -i \int_{-\infty}^{+\infty} dt_3\;  \sum_{aq}
\bigl[
    {\Sig}_{aq} ^{R}(t_1,t_3)\, {G}_{qa}^<(t_3,t_1)
  \\
  +{\Sig}_{aq}^<(t_1,t_3)\,{G}_{qa} ^{A}(t_3,t_1)
  \bigr] \,,
\end{multline}
where we introduced standardly defined\cite{Stefanucci2025}   components (advanced, retarded  and  lesser)  for  Green's functions and  correlation self-energies.  In the steady state the two-time functions depend only on the time difference, and Fourier transformation gives
\begin{equation}\label{eq:Eint_GM}
  E_{\mathrm{int}}
  = - i \int\frac{d\omega}{2\pi}\;
 \sum_{aq}
\bigl[
    {\Sig}_{aq} ^{R}(\omega)\, {G}_{qa}^<(\omega)
  \\
  +{\Sig}_{aq}^<(\omega)\,{G}_{qa} ^{A}(\omega)
  \bigr] \,,
\end{equation}
Note  ${\Sig}$ is the correlation self-energy arising from the inter-system coupling $H_{\mathrm{int}}$ (the lead self-energies are already absorbed into the bare Green's functions). 

Because there is no inter-system tunnelling between subsystems $1$ and $2$, the full Green's function is block-diagonal:
\begin{equation}
  \bm{G} = \begin{pmatrix} \bm{G}_1 & 0 \\ 0 & \bm{G}_2 \end{pmatrix}.
\end{equation}
Likewise, the self-energy is block-diagonal too. Block-diagonality holds whenever the tunnelling amplitudes $t_{k\alpha, p}$ can be brought, by a unitary rotation in lead-mode space, to a form in which each lead mode couples to at most one of the two subsystems. In this case the lead-induced self-energy  $\Sigma^{\text{tun}}_{\alpha, a b}(\omega) = 0$, regardless of whether the modes are physically grouped into one shared reservoir or several separate ones. The only intermolecular coupling is then the particle-number conserving density-density interaction. This structural condition is satisfied in most standard realisations of capacitively coupled double quantum dots and molecular junctions.

Due to the block-diagonal structure, the interaction energy becomes
\begin{multline}\label{eq:E1def}
  E_{\mathrm{int}}
  = - i \int\frac{d\omega}{2\pi}\;
 \sum_{a_1 a_2}
\bigl[
    {\Sig}_{a_1 a_2} ^{R}(\omega)\, {G}_{a_2 a_1}^<(\omega)
  \\
  +{\Sig}_{a_1 a_2}^<(\omega)\,{G}_{a_2 a_1} ^{A}(\omega)
  \bigr].
\end{multline}

Expressing retarded and advanced components in terms of lesser and greater ones
\begin{equation}
 \Sig^{R}_{a_1 a_2} (\omega)  = i  \int_{-\infty}^{+\infty}\frac{d\omega'}{2 \pi} \frac{ \Sig^>_{a_1 a_2}(\omega') -  \Sig^<_{a_1 a_2}(\omega')}{\omega -\omega' + i \eta}\,,
\label{eq:Sig_ret}
\end{equation}
\begin{equation}
\Sig^{A}_{a_1 a_2} (\omega) = \big( \Sig^{R}_{a_2 a_1}(\omega) \Big)^*,
\label{eq:Sig_adv}
\end{equation}
we get
\begin{multline}
  E_{int}
  =
\int\frac{d\omega}{2\pi} \frac{d\omega'}{2\pi} 
\\
 \sum_{a_1 a_2}\left[
  \frac{\Sig^<_{a_1 a_2}(\omega) G^>_{a_2 a_1}(\omega') -  \Sig^>_{a_1 a_2}(\omega) G^<_{a_2 a_1}(\omega') }{\omega -\omega' -i \eta}
  \right].
\end{multline}

\subsection{Second Born evaluation of the interaction energy}
\label{sec:2Born}

At second order in the inter-system Coulomb coupling $U$, the correlation self-energy for system $1$ on the Keldysh contour reads
\begin{multline}\label{eq:Sig_contour}
  \Sig_{a_1 a_2}(\tau_1,\tau_2)
  = i \sum_{b_1 b_2}
    U_{a_1 b_1}\,U_{a_2 b_2}
  \\
  \times  G_{a_1 a_2}(\tau_1,\tau_2)\;
    \Pi_{b_1 b_2}(\tau_1,\tau_2) \,,
\end{multline}
where the polarisation propagator (particle-hole bubble) of system $2$ is
\begin{equation}\label{eq:Pi_contour}
  \Pi_{b_1 b_2}(\tau_1,\tau_2)
  = -i\,G_{b_1 b_2}(\tau_1,\tau_2)\,G_{b_2 b_1}(\tau_2,\tau_1) \,.
\end{equation}
Diagrammatically, $ \Sig_{a_1 a_2}(\tau_1,\tau_2)$ consists of the Green's function line $G_{a_1 a_2}$ dressed by the fluctuation propagator $\Pi_{b_1 b_2}$ of the other subsystem, with two Coulomb vertices $U_{a_1 b_1}$ and $U_{a_2 b_2}$.

Applying the Langreth rules\cite{Haug2008} to Eq.~\eqref{eq:Pi_contour} yields the lesser and greater components in frequency space:
\begin{align}
  \Pi^<_{b_1 b_2}(\omega)
  &= -i\int\frac{d\omega'}{2\pi}\;
     G^<_{b_1 b_2}(\omega')\;
     G^>_{\,b_2 b_1}(\omega' - \omega)  \,,\label{eq:Pi_less}\\
  \Pi^>_{b_1 b_2}(\omega)
  &= -i\int\frac{d\omega'}{2\pi}\;
     G^>_{b_1 b_2}(\omega')\;
     G^<_{\,b_2 b_1}(\omega' - \omega) \,.\label{eq:Pi_great}
\end{align}
These satisfy the crossing symmetry
\begin{equation} \label{eq:Switch}
\Pi^<_{b_1 b_2}(\omega) =  \Pi^>_{\,b_2 b_1}(-\omega) \,,
\end{equation}
which follows from a change of integration variable $\omega' \to \omega' + \omega$ in Eq.~\eqref{eq:Pi_less}.

Similarly, the lesser and greater self-energies in frequency space are
\begin{multline}
  \Sig^{<,>}_{a_1 a_2}(\omega)
  = i \sum_{b_1 b_2} U_{a_1 b_1}\,U_{a_2 b_2}
  \\
  \times \int \frac{d\omega'}{2 \pi}\;
     G^{<,>}_{a_1 a_2}(\omega')\;\Pi^{<,>}_{b_1 b_2}(\omega -\omega') \,.
  \label{eq:Sig_lessgtr}
\end{multline}

Inserting the second Born self-energies~\eqref{eq:Sig_lessgtr} and collecting the Green's function products into polarisation propagators of system $1$, we arrive at
\begin{widetext}
\begin{equation}\label{eq:Eint_comp}
  E_{\mathrm{int}}
  = -\sum_{\substack{a_1 a_2 \\ b_1 b_2}} U_{a_1 b_1}\,U_{a_2 b_2} \int\frac{d\omega_1}{2\pi} \frac{d\omega_2}{2\pi}  
  \frac{ 
     \Pi^<_{a_1 a_2}(\omega_1)\;\Pi^<_{b_1 b_2}(\omega_2) - 
     \Pi^>_{a_1 a_2}(\omega_1)\;\Pi^>_{b_1 b_2}(\omega_2) }{\omega_1 +\omega_2} \,.
\end{equation}
\end{widetext}

Here, we use the crossing symmetry~\eqref{eq:Switch} to show that the numerator vanishes on the singularity surface $\omega_1 + \omega_2 = 0$, therefore  the infinitesimal $i\eta$ can be dropped from the equation.
This is the formula for the nonequilibrium dispersion interaction energy, it is determined entirely by the polarisation propagators of the individual subsystems, evaluated in their respective NESSs.

\subsection{Noise-response decomposition of dispersion energy}
\label{sec:noise_response}

The polarisation propagators $\Pi^{<,>}$ encode two distinct physical quantities: charge noise and charge dissipation. For a single-orbital subsystem (the multilevel generalisation follows straightforwardly), the density fluctuation operator is $\delta\hat{n}(t) = d^\dagger(t)\,d(t) - \langle\hat{n}\rangle$. Evaluating the two-time density-density correlator within the self-consistent Born approximation and neglecting vertex correction diagrams, one obtains
\begin{align}
  \Pi^>(\omega) &= -i\,S^+(\omega) \,,\label{eq:Pi_great_S}\\
  \Pi^<(\omega) &= -i\,S^-(\omega) \,,\label{eq:Pi_less_S}
\end{align}
where the real, non-negative spectral weights
\begin{align}
S^+(\omega)
  &= \int  dt e^{i\omega t}\,
     \langle\delta\hat{n}(t)\,\delta\hat{n}(0)\rangle \geq 0
 \,,\label{eq:Splus}\\
  S^-(\omega) 
  &= \int  dt e^{i\omega t}\,
     \langle\delta\hat{n}(0)\,\delta\hat{n}(t)\rangle \geq 0
 \,,\label{eq:Sminus}
\end{align}
are the Fourier transforms of the physical density-density correlators.

The symmetrised and antisymmetrised combinations define the charge noise spectrum and the dissipative charge response, respectively
\begin{align}
  S(\omega) &= \frac{S^+(\omega)+S^-(\omega)}{2} = \frac{i}{2}\bigl[\Pi^>(\omega)+\Pi^<(\omega)\bigr] \,,\label{eq:S_def}\\
  \chi(\omega) &=\frac{S^+(\omega)-S^-(\omega)}{2} = \frac{i}{2}\bigl[\Pi^>(\omega)-\Pi^<(\omega)\bigr] \,.\label{eq:chi_def}
\end{align}
The noise spectrum $S(\omega)$ is the spectral density of the symmetrised (anticommutator) charge correlator and represents the total charge fluctuation activity at frequency $\omega$. The dissipative response $\chi(\omega)$ is connected to the commutator of density operators. It measures the directional asymmetry between energy-absorbing and energy-releasing processes at frequency $\omega$.

The dispersion interaction energy takes the noise-response form
\begin{widetext}
\begin{equation}\label{eq:Eint_noise_resp}
  E_{\mathrm{int}}
  = -2\sum_{\substack{a_1 a_2\\b_1 b_2}}
    U_{a_1 b_1}\,U_{a_2 b_2}
    \int  \frac{d\omega_1}{2\pi}  \frac{d\omega_2}{2\pi}\;
    \frac{
      S_{a_1 a_2}(\omega_1)\;\chi_{b_1 b_2}(\omega_2)
    + \chi_{a_1 a_2}(\omega_1)\;S_{b_1 b_2}(\omega_2)
    }{\omega_1+\omega_2}\,.
\end{equation}
\end{widetext}
This result holds at arbitrary voltage and temperature. The interaction energy has the structure of a fluctuation-dissipation coupling: charge noise on system $1$ ($S_{a_1 a_2}$) couples through the Coulomb interaction $U^2$ with the dissipative charge response of system $2$ ($\chi_{b_1 b_2}$), and vice versa. System $1$ fluctuates; system $2$ absorbs (or emits); the energy denominator $(\omega_1 + \omega_2)^{-1}$ weights the correlated fluctuation-absorption processes.

In equilibrium, the fluctuation-dissipation theorem locks $S$ and $\chi$ together: $S(\omega) = \chi(\omega)\coth(\beta\omega/2)$. Out of equilibrium, this lock breaks, and the noise and response become independent. The voltage redistributes the charge fluctuation spectral weight across frequencies without respecting detailed balance, fundamentally altering the dispersion interaction.

%=====================================================================
\subsection{Equilibrium: KMS relations and universal attractiveness }
\label{sec:equilibrium}
%=====================================================================

We now prove that the dispersion interaction energy is strictly non-positive in thermal equilibrium, independent of the details of the subsystem Hamiltonians.

In equilibrium, the single-particle Green's functions satisfy the fermionic KMS condition (fluctuation-dissipation theorem),
\begin{equation}
  G^<_{bb'}(\omega) = -e^{-\beta(\omega - \mu)} G^>_{bb'}(\omega) \,.
  \label{eq:fermionic_KMS}
\end{equation}
This yields the bosonic KMS relation for the polarisation propagators
\begin{equation}
  \Pi^>_{b_1 b_2}(\omega)
  = e^{\beta\omega} \Pi^<_{b_1 b_2}(\omega) \,.
  \label{eq:bosonic_KMS}
\end{equation}

Applying Eq.~\eqref{eq:bosonic_KMS} to both subsystems, the numerator in Eq.~\eqref{eq:Eint_comp} factorises as
\begin{equation}
\Pi^<_{a_1a_2}(\omega_1)\,\Pi^<_{b_1b_2}(\omega_2)\,\bigl[1 - e^{\beta(\omega_1+\omega_2)}\bigr] \,.
\end{equation}
Using $\Pi^<(\omega) = -i\,S^-(\omega)$ with $S^-(\omega) \geq 0$, the product $\Pi^<_1\,\Pi^<_2 = -S^-_1\,S^-_2 \leq 0$. The kernel
\begin{equation}
K(\Omega) = \frac{1 - e^{\beta\Omega}}{\Omega}
\end{equation}
is strictly negative for all $\Omega \neq 0$: for $\Omega > 0$, the numerator is negative and the denominator positive; for $\Omega < 0$, the numerator is positive and the denominator negative; and $K(0) = -\beta < 0$.

Combining these signs with $U_{a_1 b_1}\,U_{a_2 b_2} \geq 0$, we obtain
\begin{equation}\label{eq:Eint_leq_0}
  E_{\mathrm{int}} \leq 0 \,.
\end{equation}
The equilibrium dispersion interaction is universally attractive. The KMS condition locks the relative weight of particle-hole excitations and de-excitations such that correlated charge fluctuations on the two subsystems always lower the energy. This result holds irrespective of the number of orbitals, the single-particle spectrum, or the strength of coupling to the reservoirs.

\subsection{Nonequilibrium: Generalised KMS relations and sign change}
\label{sec:noneqKMS}
For two identical subsystems driven by the same bias voltage $V$, the interaction energy~\eqref{eq:Eint_comp} simplifies to 
\begin{multline}
\label{eq:Eint-neq}
E_{\mathrm{int}} = -U^2 \int \frac{d\omega_1}{2\pi}\frac{d\omega_2}{2\pi}
\\
\times\frac{\Pi^<(\omega_1)\,\Pi^<(\omega_2) - \Pi^>(\omega_1)\,\Pi^>(\omega_2)}{\omega_1+\omega_2} \,,
\end{multline}
where the orbital indices for a single-level system have been dropped for transparency of equations.

Out of equilibrium, the KMS identity no longer holds. However, provided $\Pi^<(\omega) \neq 0$, one can always define a generalised KMS ratio
\begin{equation}
\label{eq:gen-KMS}
\Pi^>(\omega) = Z(\omega, V)\;\Pi^<(\omega) \,,
\end{equation}
where $Z(\omega,V)$ reduces to $e^{\beta\omega}$ in equilibrium. In terms of the spectral weights,
\begin{equation}
Z(\omega,V) = \frac{S^+(\omega)}{S^-(\omega)} \,,
\end{equation}
which is the ratio of the energy-absorbing to energy-releasing fluctuations at frequency $\omega$.  The crossing symmetry $S^+(\omega) = S^-(-\omega)$, implies the inversion identity
\begin{equation}
\label{eq:Z-inversion}
Z(\omega,V)\,Z(-\omega,V) = 1.
\end{equation}

The dispersion interaction is attractive when the charge fluctuations obey an effective fluctuation-dissipation relation with positive temperature ($Z(\omega,V)> 1$), vanishes when $Z(\omega,V)=1$, and becomes repulsive under population inversion ($ Z(\omega,V)< 1$).

%=====================================================================
\section{Model Calculations}
\label{sec:model}
%=====================================================================

\subsection{Model Hamiltonian}

We consider two identical single-level molecular junctions ($a$ and $b$), each consisting of a single electronic level coupled to left and right metallic leads. The molecular Hamiltonian is
\begin{equation}
\label{eq:Hmol}
H_1 + H_2 + H_{\mathrm{int}} = \epsilon_a\, \hat n_a + \epsilon_b\, \hat n_b + U\,(\hat n_a - q_a)(\hat n_b - q_b) \,,
\end{equation}
where $\hat n_p = d^\dagger_p d_p$ is the occupation number operator for site $p$, $\epsilon_p$ is the orbital energy and $U$ is the intermolecular Coulomb coupling. The parameters $q_a$ and $q_b$ represent the positive ionic core charges of the two molecules. Setting  $q_a = q_b = +1$, models the molecules as undergoing Highest Occupied Molecular Orbital (HOMO) dominated transport.

Each site is coupled symmetrically to left ($L$) and right ($R$) leads with the wide-band limit self-energy characterised by a constant hybridisation $\Gamma$. The bias voltage is applied symmetrically, $\mu_{L/R}^{(1),(2)} = E_F \pm eV/2$.

\subsection{Self-consistent second Born approximation calculations}

\begin{figure}[t]
\centering
\includegraphics[width=0.95\columnwidth]{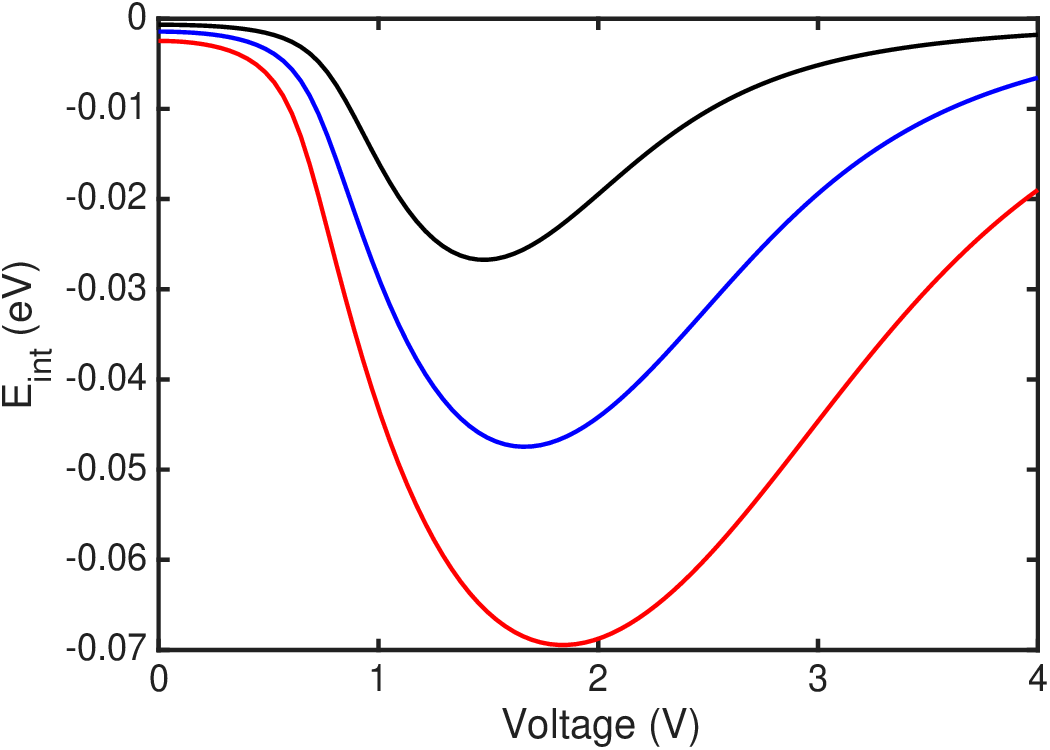}
\caption{Dispersion interaction energy $E_{\mathrm{int}}$ (in eV) as a function of applied bias voltage $V$ (in V) for two capacitively coupled single-level molecular junctions described by the Hamiltonian~\eqref{eq:Hmol}. Parameters: $\epsilon_a = \epsilon_b = -0.5$~eV, $T = 300$~K, $\Gamma = 0.05$~eV, and $q_a = q_b = +1$.  The Fermi energy set to $E_F = 0$. The three curves correspond to intermolecular Coulomb couplings: $U = 0.5$~eV (black), $U = 0.75$~eV (blue), and $U = 1.0$~eV (red). The voltage enhances the attractive dispersion interaction over an order of magnitude larger than the equilibrium values. The characteristic voltage  at which the minimum occurs shifts to higher values with increasing $U$ due to the interaction renormalisation of the effective level position. }
\label{fig:Eint_vs_V}
\end{figure}

\begin{figure}[t]
\centering
\includegraphics[width=0.95\columnwidth]{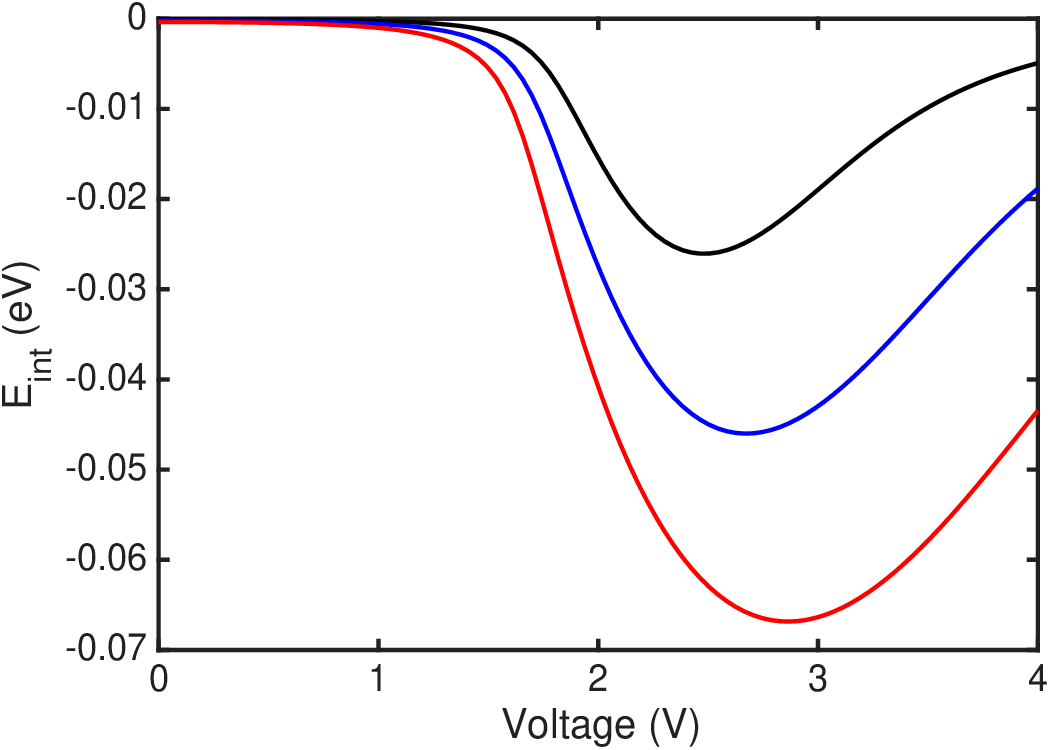}
\caption{Same as Fig.~\ref{fig:Eint_vs_V}, but for deeper molecular orbital energies $\epsilon_a = \epsilon_b = -1.0$~eV. All other parameters are identical to Fig.~\ref{fig:Eint_vs_V}. The minima shifts to substantially higher voltages compared to Fig.~\ref{fig:Eint_vs_V}, demonstrating that the position of the dispersion enhancement tracks the molecular orbital energy according to the voltage dependent pole of the retarded Green's function. The depths of the minima are nearly unchanged, indicating that the maximum attraction is set by the interplay of $U$, $\Gamma$, and the current-driven noise rather than by molecular orbital energies.}
\label{fig:Eint_vs_V_e1}
\end{figure}

The calculations are performed within the fully self-consistent second Born approximation for the intermolecular Coulomb interaction $U$. This goes beyond a perturbative treatment in which the non-interacting (or Hartree-level) Green's functions are simply substituted into the polarisation bubble diagrams; instead, the full correlation self-energies, Green's functions, and polarisation propagators are determined self-consistently.

The self-consistency cycle proceeds in two stages. In the first stage, the Hartree self-energy $\Sigma^{\R}_{\mathrm{H},p} 
= U \langle \hat{n}_{\bar{p}}\rangle $ 
\begin{equation}
\Sigma^{\R}_{\mathrm{H},p}  = -iU  \int\frac{d\omega'}{2 \pi} G^<_{\bar{p}}(\omega')\,,
\end{equation}
($\bar p$ is the opposite site to $p$) is converged to self-consistency with the lead-dressed Green's functions (i.e.\ with $\Sigma^{\R}_{\mathrm{c},p} = \Sigma^{<,>}_{\mathrm{c},p} = 0$), using linear mixing of $G^<_p$ and $G^>_p$. In the second stage, the full second Born self-energies are switched on and iterated to self-consistency on top of the converged Hartree solution. At each iteration, the polarisation bubbles~\eqref{eq:Pi_less}--\eqref{eq:Pi_great} of the partner molecule are constructed from the present $G^{<,>}_{\bar{p}}$ via trapezoidal-rule convolution, the lesser and greater correlation self-energies are assembled from Eq.~\eqref{eq:Sig_lessgtr}, and the retarded and advanced components are obtained from the Kramers-Kronig integral~\eqref{eq:Sig_ret}--\eqref{eq:Sig_adv} evaluated as a principal-value numerical integration with a small broadening $\eta$.  The retarded and advanced Green's functions are then updated via the Dyson equation, and $G^{<,>}_p$ are recalculated from the Keldysh equation using the full self-energies $\Sigma^{<,>}_{p} = \Sigma^{<,>}_{\mathrm{leads},p} + \Sigma^{<,>}_{\mathrm{c},p}$. Convergence of the second stage is accelerated by dynamic linear mixing and the iteration is terminated when the maximum change in both $G^<_p$ and $G^>_p$ across the energy grid falls below a tolerance of $10^{-10}$~eV$^{-1}$. All calculations are performed on a uniform energy grid spanning $[-10, 10]$~eV.

\subsection{Controlling dispersion interaction by applied voltage}

Figure~\ref{fig:Eint_vs_V} shows the dispersion interaction energy $E_{\mathrm{int}} $ as a function of the applied bias voltage $V$ for three values of the intermolecular Coulomb coupling $U$, with the molecular orbital energies fixed at $\epsilon_a = \epsilon_b = -0.5$~eV relative to $E_F$. The results reveal a striking nonequilibrium enhancement of the dispersion force.

At zero bias, the equilibrium dispersion interaction is weakly attractive for all three values of $U$, with magnitudes of order $10^{-3}$~eV. This small equilibrium value reflects the narrow hybridisation width ($\Gamma = 0.05$~eV) relative to the orbital resonance energy  ($\sim 0.75-0.9$~eV): the charge fluctuation spectral weight on each dot is concentrated in a narrow window around the molecular resonance, and the equilibrium noise is thermally suppressed at $k_BT \approx 0.026$~eV.

As the voltage increases from zero, the dispersion interaction becomes dramatically more attractive. Each curve reaches a pronounced minimum at a characteristic voltage $V_C$: for $U = 0.5$~eV, the minimum occurs near $V_C \approx 1.5$~V with $E_{\mathrm{int}}  \approx -0.03$~eV; for $U = 0.75$~eV, the minimum sits near $V_C \approx 1.7$~V with depth $\approx -0.05$~eV; and for $U = 1.0$~eV, the minimum occurs at $V_C \approx 1.9$~V with depth $\approx -0.07$~eV. The enhancement relative to the equilibrium value exceeds an order of magnitude in all three cases.

To demonstrate that the position of the dip is controlled by the molecular orbital energy, we repeat the calculation with the orbital energies pushed deeper to $\epsilon_a = \epsilon_b = -1.0$~eV, while keeping all other parameters identical to those of Fig.~\ref{fig:Eint_vs_V}. The results are shown in Fig.~\ref{fig:Eint_vs_V_e1}.

The qualitative shape of the curves is preserved: each $E_{\mathrm{int}}(V)$ trace exhibits a pronounced attractive minimum, recovering toward small values at high bias. However, the position of the dip shifts substantially toward larger voltages: for $U = 0.5$~eV the minimum now occurs at $V_C \approx 2.5$~V, for $U = 0.75$~eV at $V_C \approx 2.7$~V, and for $U = 1.0$~eV at $V_C \approx 2.9$~V. Comparing with the corresponding values for $\epsilon_a = \epsilon_b=-0.5$~eV, we observe that doubling the molecular orbital energy shifts $V_C$ to substantially higher values, in approximate agreement with the resonance condition (see Section~\ref{sec:mechanism}), where $eV_C/2$ corresponds to the renormalised molecular orbital which is the real part of the pole of the retarded Green's function.

The depths of the minima, by contrast, are nearly unchanged between the two cases. This near-invariance of the maximum attraction with respect to the molecular orbital energy, combined with the strong shift in $V_C$, confirms that the molecular orbital energy controls where the resonant enhancement of the dispersion interaction occurs. The magnitude of the enhancement, however, is set by the universal interplay between the intermolecular Coulomb coupling $U$, the hybridisation $\Gamma$, and the spectral weight of the current-driven charge fluctuations. In other words, the orbital energy acts as a tuning parameter that selects the operating bias voltage at which the dispersion force is maximally enhanced, without significantly affecting how strong that enhancement can become.

\subsection{Repulsive dispersion interaction under population inversion}
\label{sec:repulsive_results}

The generalised KMS relations of Section~\ref{sec:noneqKMS} predicts that the dispersion interaction reverses sign, from attractive to repulsive, 
when energy-emitting charge fluctuations dominate over energy-absorbing ones resulting in  $Z(\omega,V) <1$. Physically, this requires a population-inverted electronic distribution in the leads, with more electrons occupying states above the Fermi level than below. To test this prediction within the same self-consistent NEGF calculations used in the previous subsections, we replace the standard Fermi-Dirac lead distributions by Fermi-Dirac functions evaluated at a negative electronic temperature, $T = -300$~K. The resulting Fermi-Dirac distribution function,
remains bounded between $0$ and $1$ at every frequency, but is monotonically increasing with energy - the defining feature of an inverted distribution. All identities of the NEGF formalism hold without modification, and the self-consistent second Born iterations converge normally. The negative-temperature lead distribution should be regarded as a minimal, tractable model of a population-inverted reservoir; physical realisations of inversion (optical pumping, ultrafast voltage transients, superconducting quasiparticle injection; see Section~\ref{sec:discussion}) produce non-thermal $f(\omega)$ whose detailed shape would modify the quantitative features but not the qualitative behaviour we now describe.

\begin{figure}[t]
\centering
\includegraphics[width=0.95\columnwidth]{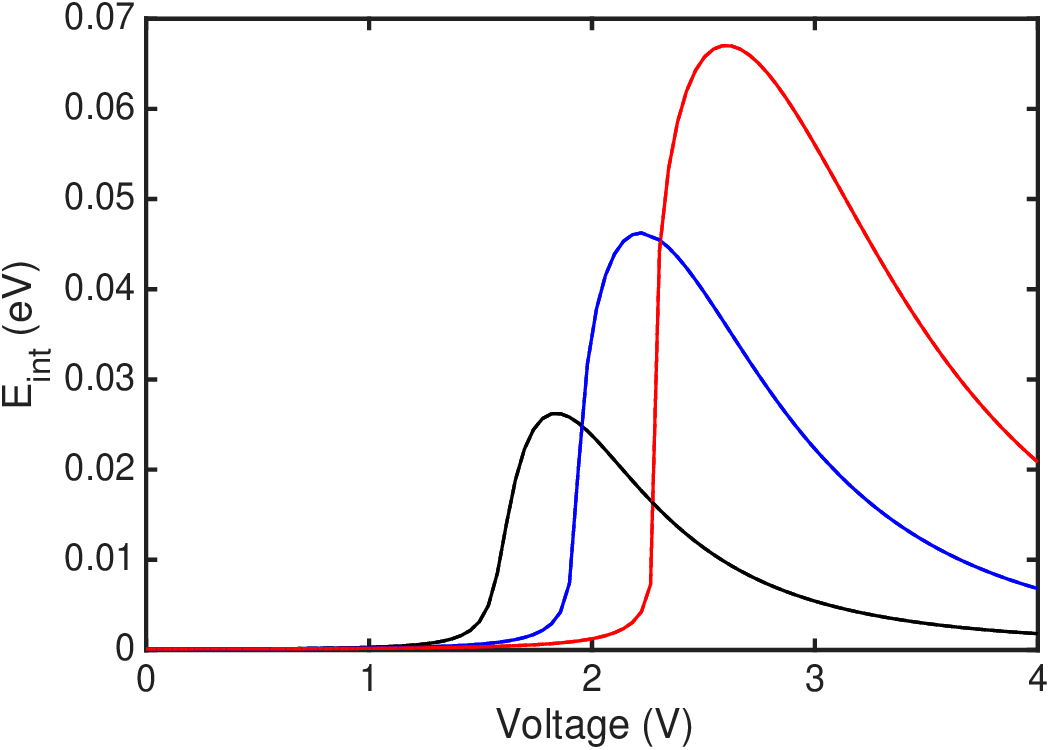}
\caption{Repulsive dispersion interaction under population inversion. The dispersion interaction energy $E_{\mathrm{int}}$ (in eV) is plotted as a function of applied bias voltage $V$ (in V) for the same parameters as Fig.~\ref{fig:Eint_vs_V}, but with the lead electronic temperature set to $T = -300$~K to model a population-inverted lead distribution. The interaction is  repulsive ($E_{\mathrm{int}}  > 0$) at every bias.}
\label{fig:Eint_vs_V_negT}
\end{figure}

Figure~\ref{fig:Eint_vs_V_negT} shows the dispersion interaction energy $E_{\mathrm{int}} $ as a function of bias voltage for the same parameters used in Fig.~\ref{fig:Eint_vs_V}, but with the lead temperature set to $T = -300$~K.
The interaction is now repulsive ($E_{\mathrm{int}} > 0$) at every voltage, with a pronounced maximum repulsion and a decay toward zero at both low and high bias. 

%=====================================================================
\section{Discussion}
\label{sec:discussion}
%=====================================================================

\subsection{Physical mechanism of the voltage enhancement}
\label{sec:mechanism}

The enhancement of the dispersion interaction at finite bias, and the location of the minimum in $E_{\mathrm{int}}(V)$, can be understood from a simple and physically transparent resonance condition involving the pole of the retarded Green's function.

In the self-consistent second Born approximation, the retarded Green's function of each molecular dot takes the form
\begin{equation}
\label{eq:GR_full}
G_p^{\R}(\omega) = \frac{1}{\omega - \epsilon_p +U q_{\bar p} - \Sigma^{\R}_{\mathrm{leads}, p}(\omega) - \Sigma^{\R}_{\mathrm{H}, p} - \Sigma^{\R}_{\mathrm{c},p}(\omega)} \,.
\end{equation}
The real part of the denominator defines the effective (renormalised) molecular orbital energy,
\begin{equation}
\label{eq:eps_eff}
\tilde{\epsilon}_p(V) = \epsilon_p-U q_{\bar p} + \mathrm{Re}\,\Sigma^{\R}_{\mathrm{leads},p} + \Sigma^{\R}_{\mathrm{H},p}(V)+ \mathrm{Re}\,\Sigma^{\R}_{\mathrm{c},p}(\tilde{\epsilon},V) \,,
\end{equation}
which incorporates the real part of the lead self-energy (vanishing in the wide-band limit),  the static Hartree shift from the intermolecular Coulomb interaction and interaction with positive ion core charge $q_{\bar p}$, and  the dynamic correlation shift from the second Born self-energy. The imaginary part of the denominator gives the spectral broadening of the pole, of order $\Gamma$, augmented by a smaller correlation-induced contribution from $\mathrm{Im}\,\Sigma^{\R}_{c}$.

The dispersion interaction energy~\eqref{eq:Eint_comp} is controlled by the polarisation propagators $\Pi^{<,>}(\omega)$, which are particle-hole bubbles built from the dressed Green's functions. Particle-hole excitations across the molecule-lead interface become maximally effective when a lead chemical potential $\mu_{L/R}$ crosses the pole of $G^{\R}$. At this crossing, the spectral function $A(\omega) = -2\,\mathrm{Im}\,G^{\R}(\omega)$ has a peak inside the bias window, the molecular orbital is strongly hybridised with the continuum of lead states, and both the occupied-state weight and the empty-state weight at the resonance become simultaneously accessible. The result is a sharp increase in the charge noise spectrum $S(\omega)$ and in the dissipative response $\chi(\omega)$, which together drive the fluctuation-dissipation coupling in Eq.~\eqref{eq:Eint_noise_resp}.

For the symmetric bias arrangement $\mu_{L/R} = \pm eV/2$, this yields the resonance condition
\begin{equation}
\label{eq:resonance}
eV_C/2 = \tilde{\epsilon}(V_C)\,.
\end{equation}
 The condition makes several features of the numerical results, such as the shift of $V_C$ with $U$ and the shift of $V_C$ with $\epsilon$, immediately transparent.
 It also explains the monotonic decrease of the magnitude of dispersion interaction at higher voltages ($V > V_C$).

\subsection{Conditions for repulsion}
\label{sec:repulsion_conditions}

In terms of the spectral weights, the generalised KMS ratio $Z(\omega,V) = S^+(\omega)/S^-(\omega)$ measures the relative rate of energy-absorbing and energy-releasing transitions at frequency $\omega$. With normal Fermi-Dirac lead distributions, irrespective of the bias voltage, one always has $S^+(\omega) > S^-(\omega)$ for $\omega > 0$, so $Z > 1$ everywhere. This reflects the fact that higher energy states are less occupied than lower energy states, and the dispersion interaction is necessarily attractive. The sign of the interaction is therefore controlled not by the magnitude of the applied bias, but by the functional form of the electronic distribution in the leads. A large voltage applied to normally populated leads changes the magnitude of the attraction, as demonstrated in Sec.~\ref{sec:model}; only a population-inverted distribution for a sufficient range of energies can produce repulsion.

Several physical routes to such an inverted lead distribution can be envisaged.

(i) \textit{Optical pumping of the leads.} Intense laser irradiation can create a transient population inversion in the conduction band of metallic or semiconducting electrodes, generating a non-thermal electron distribution  in a window of energies above the Fermi level. The lifetime of such an inverted distribution is limited by electron-electron scattering ($\sim 10-100$~fs in metals) and electron-phonon relaxation ($\sim 1$~ps),\cite{Pelter2024} so the resulting repulsive dispersion force would be observable on ultrafast timescales.

(ii)\textit{Ultrafast voltage pulses.} Step-function voltage transients faster than the electronic relaxation time create non-thermal effective distribution functions in the junction region.\cite{Jauho1994} The resulting transient distributions may depart sufficiently far from detailed balance to produce a repulsive dispersion force.

(iii) \textit{Superconducting leads with quasiparticle injection.} In superconducting junctions, quasiparticle injection can sustain non-thermal distributions above the gap on much longer timescales (microseconds to milliseconds) due to the suppressed relaxation rate,\cite{Arutyunov_2018} providing a quasi-stationary platform on which the repulsive dispersion regime could be probed.

%=====================================================================
\section{Conclusions}
\label{sec:conclusions}
%=====================================================================

We have developed a nonequilibrium Green's function theory for dispersion interactions between current-carrying nanostructures. The main results are:

(i) The nonequilibrium dispersion interaction formula~\eqref{eq:Eint_comp} expresses the dispersion interaction energy in terms of the polarisation propagators of the individual subsystems, each evaluated in its own NESS.

(ii) The noise-response decomposition~\eqref{eq:Eint_noise_resp} reveals the fluctuation-dissipation structure of the interaction: charge noise on one subsystem couples with the dissipative charge response of the other.

(iii) In thermal equilibrium, the dispersion interaction is universally attractive, a result proved using the bosonic KMS condition. Out of equilibrium, the sign of the interaction is governed by the generalised KMS relation. The voltage can enhance the attractive force by an order of magnitude. Repulsion requires population inversion in the electronic leads.

These results establish that dispersion forces between nanostructures can be actively controlled by applied voltage, opening possibilities for voltage-tunable self-assembly and nanoscale force engineering in current-carrying environments.

\begin{acknowledgments}
This work was supported by the  Queensland Government Quantum and Advanced Technologies Talent Building Program.
\end{acknowledgments}

\clearpage
\bibliography{neq_dispersion}

\end{document}